\documentclass[12pt]{article}
\usepackage{amsmath,euscript}
\usepackage{graphicx}
\usepackage{bm}
\usepackage{epsfig}

 \newcount\hour \newcount\minute
  \hour=\time \divide \hour by 60
  \minute=\time
  \newcommand{\mydate}{\ \today \ - \number\hour :\ifnum \minute<10 0\fi 
\number\minute}

\begin{document}
\newcommand{\ltap}{\stackrel{<}{_\sim}}
\newcommand{\gtap}{\stackrel{>}{_\sim}}
\begin{titlepage}
\begin{flushright}
UCTP-102-04\\
{\tt hep-ph/0405134}
\end{flushright}

\vspace{0.7cm}
\begin{center}
\Large\bf 
Polarization in $B \to VV$ Decays
\end{center}

\vspace{0.8cm}
\begin{center}
{\sc Alexander L. Kagan}\\
\vspace{0.7cm}
{\sl Department of Physics\\
University of Cincinnati\\
Cincinnati, Ohio 45221, U.S.A.}
\end{center}

\vspace{1.0cm}
\begin{abstract}
\vspace{0.2cm}\noindent
Factorizable amplitudes in $B$ decays to light vector meson pairs give a longitudinal polarization satisfying
$1- f_L   = {\cal  O}(1/m_b^2 ) $.
This remains {\it formally} true when non-factorizable 
graphs are included in QCD factorization, and is numerically realized in $B\to \rho \rho$.  
In $\Delta S=1$ decays a QCD penguin annihilation graph
can {effectively} contribute at leading power to the transverse and longitudinal amplitudes.  The observed longitudinal polarization, $f_L (B\to \phi K^*) \approx 50\%$, can therefore be accounted for in the SM.  
The ratio of transverse rates
$\Gamma_\perp / \Gamma_\parallel $ provides
a sensitive test for new right-handed currents. 
The {\it transverse} $b\to sg$ dipole operator amplitudes are highly suppressed.
CP violation measurements can therefore discriminate between new contributions to the dipole and four quark operators.  $SU(3)_F$ violation in QCD penguin amplitudes can easily be ${\cal O}(1)$, in general, due to annihilation.  Implications for $B \to \rho K^* $ polarization and New Physics searches are pointed out.
\end{abstract}
\vspace{.75cm}
\center{\it Submitted to Physics Letters B}
\vfil

\end{titlepage}

\section{Introduction: `helicity-flip' suppression}
\label{sec:intro}

Polarization in $B \to VV$ decays should be sensitive to the $V-A$ structure of
the Standard Model 
due to the power suppression associated with 
the `helicity-flip{'} of a collinear quark. 
For example, in the Standard Model the factorizable graphs for
$\bar B\to \phi K^*$ are 
due to transition operators with 
chirality structures $(\bar s b)_{V-A}  (\bar s s )_{V\mp A}$, see Figure. 1.
There are three helicity amplitudes, ${\cal \bar A}^0$, ${\cal \bar A}^-$, and ${\cal \bar A}^+$, in which both
vectors are longitudinally, negatively, and positively polarized, respectively.
In ${\cal \bar A}^- $ a collinear $s$ or $\bar s$ quark with positive helicity 
ends up in the negatively
polarized $\phi$, whereas in 
${\cal \bar A}^+$ a second quark `helicity-flip{'} is required in the form factor transition.
In the case of new right-handed currents, e.g.,  $(\bar s b)_{V+A}  (\bar s s)_{V\pm A}$, the hierarchy is inverted, with ${\cal \bar A}^+$ and ${\cal \bar A}^-$ requiring one and two `helicity-flips{'},
respectively.  

Helicity-flip suppression can be estimated
by recalling that the probability for a positive helicity free fermion
to have negative spin along some axis is given by 
$\sin^2 \theta /2 $, where $\theta$ is the angle between the axis and the momentum vector.
For a $\phi$ meson in a symmetric configuration the transverse momentum of the valence quarks is $k_\perp \sim m_\phi /2$, implying that the helicity suppression in ${\cal \bar A}_-$ is
$\sim m_\phi / m_B $.  The form factor helicity suppression
in ${\cal \bar A}_+$ should be approximately $p_T /m_b $, where $p_T$ is the transverse momentum 
of the outgoing $s$ quark.  The latter can be estimated by identifying it
with the transverse momentum of the $b$ quark.
In the `Fermi momentum{'} model of \cite{fermimom} $<p_T^2 >^{1/2}  \approx p_F /\sqrt{3}$.
Using the equivalence of this model to a particular HQET 
based shape function ansatz \cite{roman}
and for illustration taking $\bar \Lambda \approx 500$ MeV and $-\lambda_1 \approx 0.3$ GeV$^2$ 
yields $p_F \approx 400$ MeV, or a helicity suppression of $\sim 0.05$.

These simple estimates should be compared to 
naive factorization, supplemented by the large energy 
form factor relations \cite{charlesetal} (also see \cite{benekefeldmann}).
For $\bar{B} \to \phi K^{*}$, 
\begin{equation}
\label{A0pmfact}
{\cal \bar A}^0  =  i {G_F \over  \sqrt{2}} \lambda_t^{s} { \tilde{a} } A^0_{K^* \phi},~~~
{\cal \bar A}^{\mp} = i {G_F \over  \sqrt{2}}  \lambda_t^{s} { \tilde{a} } A^{\mp}_{K^* \phi}\,.
\end{equation}
The coefficient $\tilde{a}= a_3 + a_4 + a_5 - {1\over 2} (a_7 + a_9 + a_{10})$, where the
$a_i$ are the usual naive factorization coefficients, see e.g. \cite{alietal}, and $ \lambda_p^{q} = V_{pb} V_{pq}^* $.
The large energy relations imply 
\begin{equation}
A^0_{V_1 V_2 }=   f_{V_2} m_{B\,}^2  \zeta^{V_1}_\parallel ,~~~
A^-_{V_1 V_2 }=- f_{V_2} m_{V_2} m_{B\,} 2\, \zeta^{V_1}_\perp\, , 
~~~ A^+_{V_1 V_2 }=- f_{V_2} m_{V_2} m_{B\,} 2\,  \zeta^{V_1}_\perp r^{V_1}_\perp \,.
\label{Av1v2SCET}
\end{equation}
We use the sign convention $\langle V | \bar{q} \gamma_\mu q  | 0 \rangle =-i f_V m_V
\epsilon^*_\mu$.
$\zeta_{\parallel }$ and $\zeta_{\perp}$ are the $B \rightarrow V$ form factors 
in the large energy limit \cite{charlesetal}.  Both scale as $m_b^{-3/2}$ in the heavy quark limit,
implying that helicity suppression in ${\cal\bar  A}^-  $ is 
$\approx m_\phi / m_B$ which is consistent with our estimate 
(the form factor transition contributes  
 $2\, \zeta_\perp$ in $A^-_{V_1 V_2 }$).
$r_\perp$ parametrizes the form factor helicity suppression. 
It is given by
\begin{equation}
\label{rperp}
 r_\perp = \frac{(1+m_{V_1}/{m_B} ) 
{ A^{V_1}_1}  - (1- {m_{V_1}}/{m_B} ) { V^{V_1}}     }
{(1+{m_{V_1}}/{m_B} ) 
{ A^{V_1}_1}  + (1- {m_{V_1}}/{m_B} ) { V^{V_1}}    }\,,
\end{equation}
where $A_{1,2} $ and $V$ are the axial-vector and vector 
current form factors, respectively. 
The large energy relations imply that it vanishes at leading power, because helicity suppression is ${\cal O}(1/m)$.
Light-cone QCD sum rules  \cite{ballbraun}, and lattice form factor determinations scaled to low $q^2$ using 
the sum rule approach \cite{lattice1}, give 
$r_\perp^{K^*} \approx 1-3 \,\%$; QCD sum rules give $r^{K^*}_\perp \approx 5\,\%$ \cite{fazio}; and the BSW model gives $r^{K^*}_\perp \approx 10\%$ \cite{BSW}.
These results are consistent with our simple estimate for form factor helicity suppression.

\begin{figure}
\centerline{
\hbox{
\includegraphics[width=3.3truecm,height=2.3truecm]{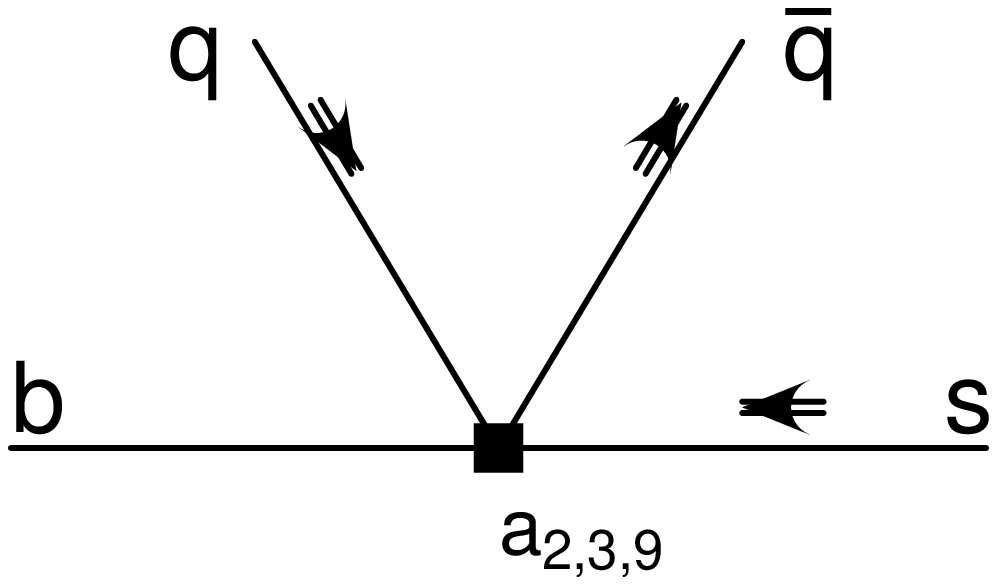}\hspace{0.85cm} 
\includegraphics[width=3.3truecm,height=2.3truecm]{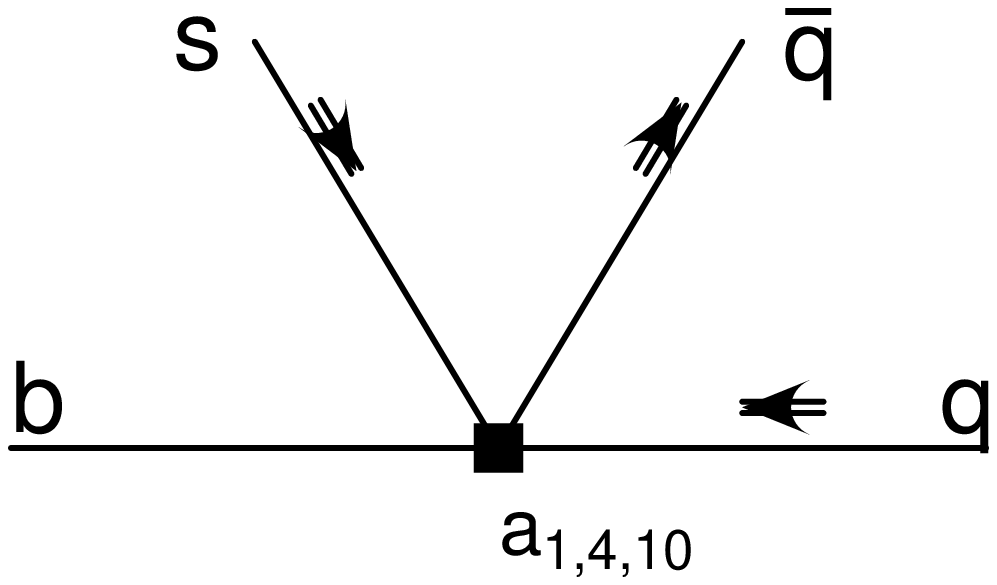}\hspace{0.85cm}
\includegraphics[width=3.3truecm,height=2.3truecm]{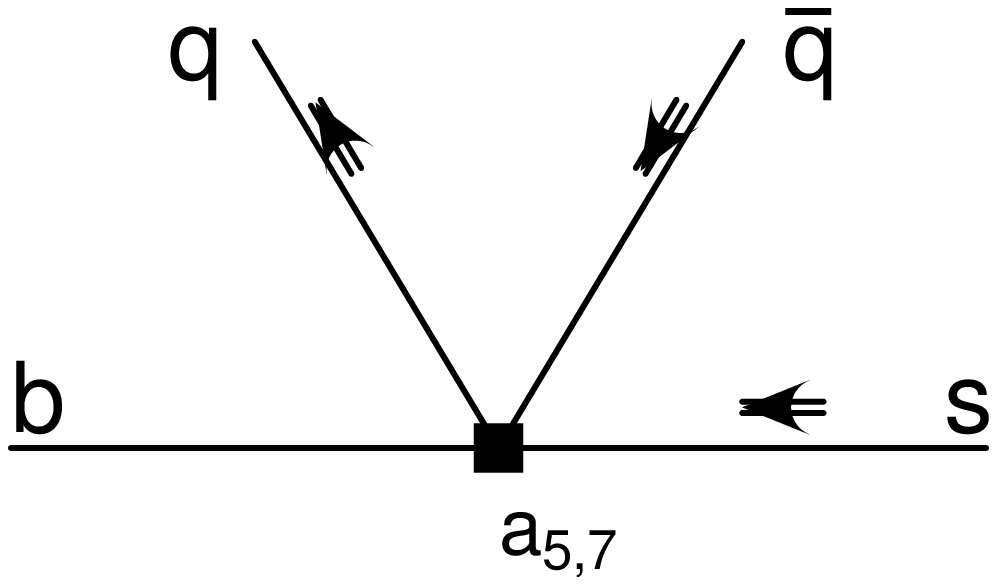}}}
\vskip -.35cm
{\caption[1]{\label{fig:factorizable}
Quark helicities (short arrows) for 
the Standard Model naive factorization coefficients $a_i$.
Upward lines form the emission vector $V_2$. $q=s$ for $\bar B \to \phi K^*$. }}
\vskip -0.2cm
\end{figure}


The large energy relations giving rise to (\ref{Av1v2SCET}) are strictly valid for the {\it soft} parts of the form factors,
at leading power and at leading order in $\alpha_s$.
However, the soft form factors are not significantly Sudakov suppressed in the Soft Collinear
Effective Theory (SCET) \cite{langeneubert}.
The results of \cite{benekefeldmann,hillFF} thus imply that 
the ${\cal O}(\alpha_s )$ form factor contributions, particularly
the symmetry breaking corrections to the large energy relations, can be neglected.
In fact, $r_\perp$ does not receive any perturbative corrections {\it at leading power}
\cite{burdmannhiller,benekefeldmann,hillFF};  again, this is because form factor helicity suppression is ${\cal O}(1/m)$.
Furthermore, power corrections to all of the form factor relations begin at ${\cal O}(1/m )$ (rather than $1/\sqrt{m}$) in SCET \cite{benekefeldmann2}.
Therefore, the above discussion of helicity suppression in naive factorization will not be significantly modified by perturbative and power corrections to the form factors.

In the transversity basis \cite{digheetal} the transvese amplitudes are
${\cal \bar A}_{\perp ,\parallel} = ( {\cal \bar A}^{-} \mp {\cal \bar A}^+)/\sqrt{2} $
(${\cal A}_{\perp ,\parallel} = ( {\cal A}^{+} \mp {\cal A}^-)/\sqrt{2} $ )
for $\bar B$ ($B$) decays.  The polarization fractions satisfy
\begin{equation}
 \label{SMpred}
1-f_L = {\cal  O}\left({1 / m_b^2}\right),
~~~
{f_\perp/ f_\parallel  }=  1+ {\cal  O} \left({1/ m_b}\right),
\end{equation} 
in naive factorization, where the subscript $L$ refers to longitudinal polarization,  $f_i = \Gamma_i /\Gamma_{\rm total}$, and $f_L + f_\perp+f_\parallel =1$.
The measured longitudinal fractions for $B \to \rho \rho$ 
are close to 1 \cite{BelleRhopRho0,BaBarRhopRho0,BaBarRhopRhom}.
This is clearly not the case for $B \rightarrow \phi K^{*0}$, for which full angular analyses
yield 
\begin{eqnarray}
\label{fLperpphiKst}
f_L &=& .43 \pm .09 \pm .04,~~~f_\perp = .41 \pm.10\pm .04~~\mbox{\cite{BellePhiKst}}\\
f_L &=& .52 \pm .07 \pm .02,~~~f_\perp = .27 \pm.07 \pm .02~~\mbox{\cite{BaBarPhiKst}}.
\end{eqnarray}
Naively averaging the Belle and BaBar measurements (without taking large correlations into account)
yields 
$f_\perp /f_\parallel = 1.4 \pm .7$. 
In the charged mode,
BaBar has measured $f_L (\phi K^{*+}) =0.46 \pm 0.12\pm 0.03$ \cite{BaBarRhopRho0}.
We must go beyond naive factorization in order to determine if the small values of $f_L (\phi K^* )$ could be due to the dominance of QCD penguin operators in $\Delta S=1$ decays.

\section{QCD factorization for $B \to VV$ decays}

In QCD factorization \cite{BBNS} exclusive two-body decay amplitudes are given in terms 
of convolutions of hard scattering kernels with meson light-cone distribution amplitudes.
At leading power this leads to factorization of short and long-distance physics.
This factorization breaks down at sub-leading powers
with the appearance of logarithmic infrared divergences.  
Nevertheless, the power-counting for all amplitudes can be obtained.  The extent to which it holds numerically can 
be checked by introducing an infrared hadronic scale cutoff,
and assigning large uncertainties. 
Non-perturbative quantities are thus roughly estimated via single gluon exchange. 
In general, large uncertainties should be expected for polarization predictions, given that
the transverse amplitudes begin at ${\cal O}(1/m)$.  However, we will find
that this is not the case for certain polarization observables, particularly after 
experimental constraints, e.g., total rate or total transverse rate, are imposed.
Our results differ substantially from previous studies of $B \to VV$ in QCD factorization \cite{chengtseng,BsVV}.  Of particular note is the inclusion of  annihilation topologies.
The complete expressions for the helicity amplitudes are lengthy and will be given in 
\cite{prd}.  Expressions for a few contributions are included below.

In QCD factorization, the Standard Model effective Hamiltonian matrix elements
can be written as  \cite{BBNS3,benekeneubert}
\begin{equation}\label{Top}
   \langle V^h_1 V^h_2 |{\cal H}_{\rm eff}|\bar B\rangle
  \! =\! \frac{G_F}{\sqrt2}\!\! \sum_{p=u,c} \! \lambda^{D}_p\,
\!   \langle V^h_1 V^h_2 |\!{\cal T_A}^{h,p}\!+\!{\cal T_B}^{h}\!|\bar B\rangle \,,
\end{equation}
where $h $ labels the vector meson helicity, and $D=s (d)$ for $\Delta S=1 (0)$.  ${\cal T_B}$ gives rise to annihilation topoplogy 
amplitudes, to be discussed shortly, and 
\begin{equation}
\langle  V^h_1 V^h_2 | {\cal T_A}^{p,h} |\bar B\rangle\! =\!
\! \sum_{i=1}^{10}\! \! a_i^{h,\,p} (V_1^\prime V_2^\prime )\langle V_1^\prime V_2^\prime   |j^i_1\!\otimes \!j^i_2| \bar B\rangle,
\end{equation}
where  $p=u,c$, and $V^\prime_1 V^\prime_2  = V_1 V_2$ or $V_2 V_1$.
The coefficients $a^{p,h}_i$ contain contributions from naive factorization, 
vertex corrections, penguin contractions, and 
hard spectator interactions. 
The transition operators $j_1^i \otimes j_2^i $ are
$\delta_{pu}    (\bar u b)_{V-A} \otimes (\bar D u)_{V-A}$ (i=1);
$\delta_{pu}   (\bar D b)_{V-A} \otimes (\bar u u)_{V-A}$ (i=2);
$[(\bar D b)_{V-A} \otimes (\bar q q)_{V\mp A} $ (i=3,5)]$\,\times \frac32 e_q $ (i=9,7);
$[ (\bar q b)_{V-A} \otimes (\bar D q)_{V-A} $ (i=4)]$\,\times \frac32 e_q $ (i=10);
$ [(\bar q b)_{S-P} \otimes (\bar D q)_{S+P} $ (i=6)]$\,\times \frac32 e_q $ (i=8), where $q$ is 
summed over $u,d,s$.
For i=1,2; 3-6; and 7-10 they originate from the current-current $Q_{1,2}$; QCD penguin
$Q_{3,..,6}$;
and electroweak penguin operators $Q_{7,..,10}$, respectively.
For $i\ne 6,8$, $\langle  V^{\prime h}_1 V^{\prime h}_2  |j^i_1 \otimes j^i_2|\! \bar B\rangle$ is defined at leading-order as 
 \begin{equation}
\langle  V^{\prime h}_1 |j^i_1|\bar B\rangle  \langle V^{\prime  h}_2  |j^i_2|0\rangle
= -i  c^i (V_1^\prime V_2^\prime ) \, A^h_{V^\prime_1 V^\prime_2 }.
\label{factdef}
\end{equation}
The $c$ coefficients contain factors of $ \pm  1$, $\pm 1/\sqrt{2}$, arising from
the vector meson flavor structures. 
$V^\prime_2$ ($V_1^\prime$) is  the `emission{'} 
(`form factor{'}) vector  meson, see Figure 1.
The $i=6,8$ matrix elements vanish at tree-level, i.e., at leading order in $\alpha_s$, as
local scalar current vacuum-to-vector matrix elements vanish.
Due to the underlying flavor structure, the effects of $a_3$-$a_{10} $ are describable in terms of a reduced set of coefficients \cite{benekeneubert}
\begin{equation}
\label{alphai}
\alpha^h_{3 (3 \rm{EW})}= a ^h_{3 (9) }+{a}^h_{5 (7)} ,~~
\alpha_{4 (4\rm{EW})}^{p , \,h}  =a_{4 (10)}^{p,\,h}  + a_{6 (8)}^{\prime p,\,h}\,,
\end{equation}
where ${a} _{6 (8)}^{\prime  p,\,h} = i {a}_{6 (8)}^{p,\,h} \langle j_1^{6 (8) } 
\!\otimes j_2^{6 (8) }\rangle  /(c^{4 (10)}  A^h_{V_1 V_2} $),
and 
$\langle j_1^{6 ,8 } \otimes  j_2^{6 ,8) }\rangle $ are 
next-to-leading order matrix elements in $\alpha_s$, in  which $ j_2$ again forms the emission
particle $V_2$.
The arguments $(V_1 V_2)$ are understood throughout.

At next-to-leading order, the  coefficients $ a_{i^{}}^{(\prime)\, p,h} $ can be written as
\cite{benekeneubert}
\begin{equation}
\label{aigeneral}
   a_{i^{}}^{(\prime)\,p,h} (V_1 V_2) = \left( C_i  \!+\! \frac{C_{i\pm 1}}{N_c}\right)
   N_i +
 C_{i\pm 1}\,\frac{{\alpha}_s  \,C_F}{ 4\pi N_c}
  \left[ V^h_i  (V_2)+\!\! \frac{4\pi^2}{N_c} \! H^h_i  (V_1 V_2) \,\right] 
+ P_i^{h,p} ( V_2)\,,
\end{equation}
where the upper (lower) signs apply when $i$ is odd (even). 
The superscript `p{'} appears for $i=4,6,8,10$.
The $N_i$ are tree-level naive factorization coefficients (Figure 1); at next-to-leading order 
the $V^h_i $ account for one-loop vertex corrections,  the $P_i^{p,h} $ for 
penguin contractions (Figure 2), and 
 the $H^h_i $ for hard spectator interactions (Figure 3).  
They are given in terms of convolutions of hard scattering kernels with vector meson and $B$ meson light-cone distribution amplitudes.  For each $i$, the corresponding graphs have the same quark helicity structure.
 

Two twist-2 light-cone distribution amplitudes
$\phi_{\parallel}  (u)$ and $\phi_{\perp}  (u)$, and four two-particle twist-3
distributions (and their derivatives) enter the longitudinal and tranverse vector meson projections \cite{balletal}. 
The argument $u $ ($\bar u \equiv  1-u$) is the quark (antiquark) light-cone momentum fraction.
The two-particle twist-3 distributions can be expressed in terms of $\phi_{\parallel,\, \perp}  (u)$ 
via Wandura-Wilzcek type equations of motion \cite{balletal}, if higher Fock states 
are ignored.   
The twist-3 vector meson projections then depend on the three distributions,
\begin{equation}
\label{WW2}
\Phi_a (u) \equiv \int_u^1 \!\!dv\,
   \frac{\phi_\parallel(v)}{v} ,~~  \Phi_b (u)\equiv  \int_0^u \!\! dv\,
    \frac{\phi_\parallel(v)}{\bar v}\,,~~ \Phi_v(u) \equiv \int_0^u\!\!dv\,\frac{\phi_\perp(v)}{\bar v}
   - \int_u^1\!\!dv\,\frac{\phi_\perp(v)}{v}\,.
\end{equation}
$\Phi_a$ and $\Phi_b$
project onto transversely polarized vectors in which
the quark and antiquark flips helicity, respectively.
$ \Phi_v(u) $, defined in  \cite{benekeneubert},
projects onto longitudinally polarized vectors in which either the quark or antiquark flips
helicity. 
Light quark mass effects are ignored, and 
a discussion of twist-4 distribution amplitudes and higher Fock state effects is deferred \cite{prd}.  
The leading-twist distribution amplitudes are given in terms of an 
expansion in Gegenbauer polynomials \cite{balletal,ballbraun}, 
\begin{equation}\label{gegenbauer}
   \phi_i (u,\mu) = 6u\bar{u}\,\bigg[ 1 + \sum_{n=1}^\infty 
   \alpha_{n,\,i} (\mu)\,C_n^{(3/2)}(2u-1) \bigg] ,\qquad i=\perp, \parallel\,.
\end{equation}
Our numerical results include the first two moments  $\alpha_{1,\,i}$, $\alpha_{2,\,i}$.
The asymptotic forms of the twist-3 distribution amplitudes are $3 \bar u^2 $, $3u^2$, and $3 (u -\bar u)$ for $\Phi_a (u)$, $\Phi_b (u) $, and $\Phi_v (u) $, respectively.  
The $B$ light-cone distribution amplitude $\phi_+^B $ \cite{benekefeldmann} enters the hard spectator interactions
through its inverse moment, $\int dl\,   {\phi^B_+ }/\,{l} = {m_B}/{\lambda_B }$. 

\begin{figure}[t]
\centerline{
\hbox{\hspace{0.2cm}
\includegraphics[width=4.5truecm,height=2.8truecm]{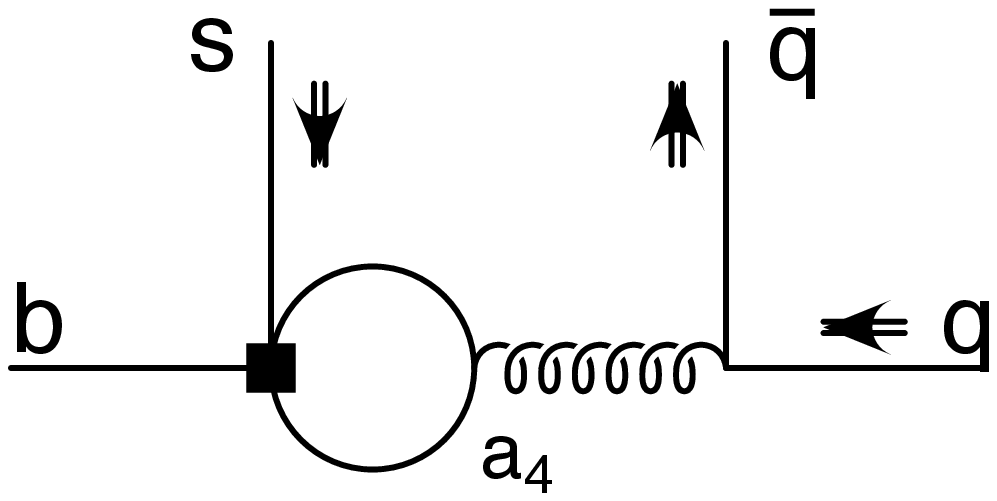}\hspace{1.5cm} 
\includegraphics[width=4.5truecm,height=2.8truecm]{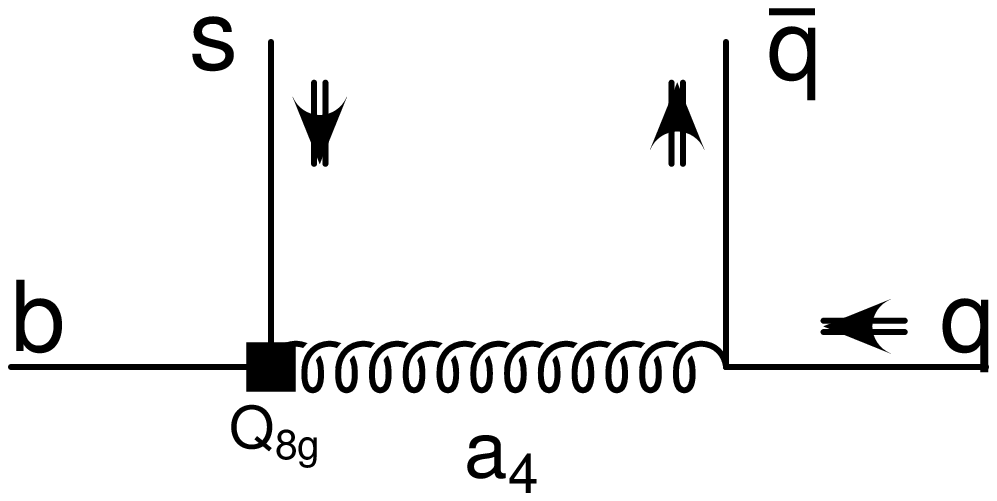}}}
\vskip -.35cm
{\caption[1]{\label{fig:penguin}
Quark helicities for $P^{p,h}_4$: 
$Q_{1,3,4,6}$ (left),  $Q_{8g}$ (right).
For $P^{p,h}_6$ flip the $q,\bar q$ helicites. Upward lines form $V_2$.}}
\vskip -0.2cm
\end{figure}

The naive-factorization coefficients are $N_i \!=\!  0$ (i=6,8), $N_i \!=\! 1$ (i$\ne $6,8 ).
The longitudinal (transverse) helicity amplitudes arise at twist-2 (twist-3) in naive facotrization.
The vertex corrections are negligible compared to the theoretical uncertainties.
Note that $V^h_{6,8} \ne 0$ for all $h$. 
At ${\cal O}(\alpha_s ,\,\alpha)$ the longitudinal penguin contractions are, respectively,
$P_{4, \,10}^{0,\,p}(V) \!=\!  P_{4, \,10} (V)$ at twist-2 and 
$P_{6 ,\,8}^{0,\,p} (V) \!=\!-\,P_{6,\,8} (V) \,2\, m_{V_2} f^\perp_V (\mu) / m_b (\mu ) f_V $ at twist-3.
The quantities $P_{i} (V)$ are the $VP$ counterparts defined in \cite{benekeneubert},
and $f_V^\perp$ is the scale-dependent tensor-current decay constant.
The transverse penguin contractions are $P_{6,\,8}^{\pm,\, p} =0$ to twist-4, and at twist-3, 
\begin{eqnarray}\label{PC4pm}
P_4^{\pm,\, p}(V_2)&\!=\!&({{\alpha}_s  \,C_F}/{ 4\pi N_c} )(
    C_1 [L - G^\pm_{V_2}(s_p) ]\!
    + C_3 [ 2 L - G^\pm_{V_2}(0) \!-\!  G^\pm_{V_2}(1) ]\nonumber\\
&+&(C_4\!+\!C_6)
   [\tilde{L}\!
    - \!3\,G^\pm_{V_2}(0)\! -\! G^\pm_{V_2}(s_c)\! -\! G^\pm_{V_2}(1) ]\,),\nonumber \\
P_{10}^{\pm,\, p}(V_2)&\!=\!& ({\alpha}/9\pi)\,
     (C_1 /N_c  +  C_2 ) [L- G^\pm_{V_2}(s_p) ],
   \end{eqnarray}
where $L= (4/3) \ln {m_b}/{\mu} + 2/3$,  $\,\tilde{L} =(20/3) \ln {m_b}/{\mu}$, $s_{p}= m_p^2 /m_b^2$ ($s_u=0$, $s_c = m_c^2 /m_b^2$), 
\begin{equation}
\label{gfcn}
   G^\pm_{V_2} (s) =  \int_0^1\!du\,G(s-i\epsilon,1-u)\, \Phi_{a,b}^{V_2} (u) \,,
      \end{equation}
and $G(s,x)$ is the well known penguin function, see e.g.,  \cite{BBNS3}.
The penguin contractions account for approximately
30\% and 20\% of the magnitudes of $\alpha_4^{c,\,0}$ and $\alpha_4^{c\,-}$
(for default input parameters), respectively, before including the hard spectator interactions.

The dipole operators $Q_{8g}$,  $Q_{7\gamma}$ do not contribute to the
{\it transverse} penguin amplitudes at ${\cal O}(\alpha_s)$ due to angular momentum conservation:  the dipole tensor current couples to a transverse gluon, but a `helicity-flip' for $q$ or $\bar q$ 
in Figure 2 would require a longitudinal gluon coupling.
Formally, this result follows from the Wandura-Wilczek relations 
and the large energy relations between the tensor-current and vector-current form factors \cite{prd}.  For example, the integrand of the convolution integral for $P_4^-$ vanishes 
identically, whereas $P_4^+ \propto \int \!\! du\, (\Phi_a^{V_2} (u) - \Phi_b^{V_2} (u) ) =0$.
Note that  transverse amplitudes in which a vector meson contains a collinear higher Fock state gluon also vanish at ${\cal O}(\alpha_s)$ . 
This can be seen from 
the vanishing of the corresponding partonic dipole operator graphs in the same momentum configurations.
Transverse ${\cal O}(\alpha_s^2 )$ spectator interaction contributions are 
highly suppressed and are studied in \cite{prd}.

The hard spectator interaction quantities $H_i^h$ contain 
logarithmically divergent integrals beyond twist-2,
corresponding to the soft spectator limit in $V_1$, see e.g., Figure 3. 
We integrate the quark light-cone momentum fraction in $V_1$
over the range $ [0,1-\varepsilon]$, and replace the divergent quantities $\ln \varepsilon$
with complex parameters $X_{H}$.  As in \cite{BBNS3,benekeneubert},
these are modeled as $X_H  = (1 \!+\! \varrho_H e^{i \varphi_H} )   \ln{m_B}/{\Lambda_h}$,
with $ \varrho_H \ltap 1$ and $\Lambda_h\approx 0.5$ GeV.  This reflects the physical ${\cal O}(  \Lambda_{\rm{QCD}})$ infrared cutoff, and allows for large strong phases from soft rescattering.
For $i \!\ne \!6,8$: $H_i^0$ first arises at twist-2; $H_i^-$ arises via a twist-3$^{V_2} \times \!$
twist-2$^{V_1}$ projection; and $H_i^+$ arises via a (twist-3)$^2$ projection. 
For $i\!=\!6,\,8$: $H_{6,\,8}^{0}\! =\!0 $ to twist-3; $H_{6,\,8}^- $ arises via a twist-3$^{V_1} \times \!$
twist-2$^{V_2}$ projection and is infrared finite; and 
$H_{6,\,8}^+$ arises via a twist-4$^{V_2}\times \!$ twist-2$^{V_1}$ projection.



The basic building blocks for annihilation are matrix elements of the operators
$(\bar q_1 b)_{V-A}  (\bar{q}_2 q_3 )_{V-A} $, $(\bar q_1 b)_{V-A}  (\bar{q}_2 q_3 )_{V+A} $, and $(\bar q_1 b)_{S-P}  (\bar{q}_2 q_3 )_{S+P} $, denoted 
$A_{1}^{i,f\, (h)} (V_1 V_2)$, $A_{2}^{i,f\,(h)} (V_1 V_2)$, and $A_{3}^{i,f\,(h)} (V_1 V_2)$,
respectively, see e.g., Figure 3.  The first quark bilinear corresponds to the $\bar B$ meson,
the superscript $i \,(f)$ indicates a gluon attached to the initial (final) state quarks 
in the weak vertex, and by convention $V_2$ ($V_1$) contains a quark (antiquark) from the weak vertex.  
$A_3^{f,0}$ and $A_3^{f,-}$ dominate the
$\Delta S=1$ QCD penguin annihilation amplitudes.
The latter are expressed as
$\langle V^h_1 V^h_2 | {\cal T}_B^{p,\,h} |\bar  B \rangle_{\rm QCD} = -i  c(V_1 V_2 ) f_B f_{V_1} f_{V_2} \,b_3^h (V_1 V_2 )$, where 
\begin{equation}
\label{b3}
b_3^h =\frac{C_F}{N_c^2} \Big[ C_3 A_1^{i,h} \!+(C_3 +N_c C_4 )A_1^{f,h}
  \!  +C_5 A_3^{i,h} \!+ (C_5 + N_c C_6 ) A_3^{f,h} \Big]\,.
 \end{equation}
The arguments $(V_1 V_2)$ have been suppressed.  The $c$ coefficients are again determined by the vector meson flavor structures.
For the electroweak penguin annihilation amplitude $b^h_{3\,EW} $, substitute $C_{3,4,5,6}\to 
 C_{9,10,7,8}$, respectively. 
$A_3^{f,0}$ and $A_3^{f,-}$ arise at twist-3.  $A_3^{f,\,-}$ is given by
\begin{equation}
\label{A3f}
 A_3^{f,\,-} \!=\! -\pi\alpha_s \int_0^1\! du \,dv\,
\Big(\,\frac{2 m_{V_2} f_{V_1}^\perp }{m_b f_{V_1}}\,\phi^{V_1}_{\perp}(v)\,\Phi^{V_2}_{b} (u)\,
 \frac{2}{v^2 \bar u} 
+\,  \frac{2 m_{V_1} f_{V_2}^\perp }{m_b f_{V_2}}\,\phi^{V_2}_{\perp}(u)\,\Phi^{V_1}_{a}(v)\,
\frac{2}{\bar u^2 v}   \Big)\,.
 \end{equation}
For $A_3^{f,\,0}$ substitute $f^{(\perp)}_{V_1} \leftrightarrow f^{(\perp)}_{V_2}$,
$\phi_\perp \to \phi_\parallel $, $\Phi_{b,a} \to \Phi_v$, and change the sign of the second term.
The integrals over $u$ and $v$ are logarithmically divergent, corresponding to the soft gluon limit
$\bar u , v \to 0$.  For simplicity, the asymptotic distribution amplitudes
are used, as in \cite{BBNS3,benekeneubert}; non-asymptotic $SU(3)_F$ violating 
effects will be discussed shortly.
The logarithmic divergences are again replaced with complex parameters, 
$X_A  = (1 \!+\! \varrho_A \,e^{i \varphi^{}_A} )   \ln\, {m_B}/{\Lambda_h}$, yielding
\begin{equation}
\label{XAA3}
 A_3^{f,\,-} \approx \pi\alpha_s  \,18\, \bigg(  \frac{2 m_{V_2} f_{V_1}^\perp }{m_b f_{V_1}}+ \frac{2 m_{V_1} f_{V_2}^\perp }{m_b f_{V_2}} \bigg)   ( 2 X^-_A -3)(1-X^-_A).
 \end{equation}
For $A_3^{f,\,0}$ substitute $f^{(\perp)}_{V_1} \leftrightarrow f^{(\perp)}_{V_2}$ and
$(2 X^-_A -3 ) (1-X^-_A ) \to (2 X^0_A -1 ) (2-X^0_A )$.  
The contributions of $ A_3^{f,\,- }$ and $ A_3^{f,\,0}$ to the helicity amplitudes  
are formally of ${\cal O}(1/m^2 )$ ($f_B$ scales like $m_b^{-1/2}$). 
However, note that as $\rho_A^{-  }$ and $\rho_A^{0 }$ are varied from 0 to 1,
$A_3^{f,\,-}$ and $A_3^{f,\,0}$ increase by more than an {\it order of magnitude}.

A summary of power counting at next-to-leading order is given in \cite{prd,talks}.
As expected, each quark `helicity-flip{'} costs $1/m$ in association with either one unit of twist,
or form factor suppression.   
A $\pm1$ change in vector meson helicity due to a collinear gluon in a higher Fock state
also costs one unit of twist, or $1/m$.
In addition, annihilaton graphs receive an overall $1/m$ suppression.
An apparent exception is provided by the (twist-3)$^2$ contributions to $A_{1,2}^{i,\, -}$; they contain
a linear infrared divergence which would break the power counting.  ($A_{1,2}^{i,\, -}$ would be promoted to ${\cal O}(1/m^2)$ but
would remain numerically small, as can been by parametrizing
the divergence as $(m_B /\Lambda_h )  \kappa e^{i \varphi} $ with, e.g., $\kappa \ltap 3$).  However,
the divergence should be canceled 
by twist-4$\times$twist-2 effects, see below.
Regardless, (\ref{SMpred}) remains formally true in QCD factorization.
The first relation in (\ref{SMpred}) has also been confirmed recently in 
SCET \cite{ianetal}.  We expect that the 
power counting obtained in QCD factorization will be reproduced for all
corresponding graphs in SCET.

Amplitudes involving twist-4 vector meson
projections remain to be explicitly evaluated \cite{prd}.
Twist-4$\times $twist-2 projections give rise to 
$H_{6,8}^+$. However, these effects should be similar in magnitude to  
(twist-3)$^2$ contributions to the positive helicity hard spectator amplitudes, which were found to be small.
The twist-4$\times$twist-2 contributions to $A_{1,2}^{f \pm}$
must cancel the non-vanishing (twist-3)$^2$ 
contributions, since $A_{1,2}^{f \pm}$ must vanish by equations of motion.  This condition leads to new Wandura-Wilczek type relations
between the products of twist-4$\times $twist-2 and (twist-3)$^2$ light-cone distribution amplitudes.  These relations should insure cancelation of the aforementioned linear divergence in 
$A_{1,2}^{i -}$ by twist-4$\times $twist-2 effects \cite{prd}.  
Finally, twist-4$\times$twist-3 projections
give rise to $A_{3}^{i,\,+}$ and $A_{3}^{f,\,+}$; however, 
these amplitudes should be both formally and numerically suppressed by ${\cal O}(1/m^2)$ compared to $A_3^{i,\,-}$ and $A_3^{f,\,-}$, respectively. 
We have also not explicitly considered
graphs in which 
higher twist two-body vector meson projections are replaced with higher Fock-state projections of same twist containing collinear gluons, e.g., $\bar q q g$.
The latter are expected to receive additional suppression at each twist, e.g., 20\% \cite{pirjolwyler}.  These corrections, especially the tree-level twist-3 contributions to the coefficients $\alpha_{i}^{(p),\,-}$, should be included \cite{prd}.
However, they will not alter our conclusions,
given the large uncertainties that have already been assigned to the power corrections. 

Expressions for a few $\bar B\to VV $ amplitudes are given below, 
\begin{eqnarray}
\label{VVamplitudes}
{\cal A}^h_{\bar{B}^0 \rightarrow  K^{*0} \phi }& =& -i {G_F \over \sqrt{2}}
\sum_{p=u,c} \lambda_p^{(s)} A_{K^* \phi }^h \left[ \alpha_3^h + \alpha_4^{p,\,h} 
-\frac12 \alpha_{3,\,{\rm EW}}^h -\frac12 \alpha_{4,\,{\rm EW}}^{p,\,h} + \beta_3^h
-\frac12 \beta_{3,\,{\rm EW}}^h  \right]\,, \nonumber\\
{\cal A}^h_{B^-\rightarrow  \rho^{-} \bar{K}^{*0} }& =& -i {G_F \over \sqrt{2}}
\sum_{p=u,c} \lambda_p^{(s)} A_{\rho K^* }^h \left[  \alpha_4^{p,\,h} 
-\frac12 \alpha_{4,\,{\rm EW}}^{p,\,h} + \delta_{pu} \beta_2^h + \beta_3^h
+\beta_{3,\,{\rm EW}}^h  \right]\,, \nonumber\\
{\cal A}^h_{\bar{B}^0\rightarrow  \rho^{+} \rho^{-} }& =& -i {G_F \over \sqrt{2}}
\sum_{p=u,c} \lambda_p^{(d)} A_{\rho \rho }^h \left[ \delta_{pu} a_1^h \!+\!
 \alpha_4^{p,\,h} 
\!+\! \alpha_{4,\,{\rm EW}}^{p,\,h}\! +\! \delta_{pu} \beta_1^h \!+\! \beta_3^h
\!+\! 2 \beta_4^h\! -\!\frac12 \beta_{3,\,{\rm EW}}^h \!  +\!\frac12 \beta_{4,\,{\rm EW}}^{p,\,h}
 \right], \nonumber\\
{\cal A}^h_{{B}^- \rightarrow  \rho^- \rho^{0}}& =& -i {G_F \over {2}}
\sum_{p=u,c} \lambda_p^{(d)} A_{\rho \rho }^h \left[ \delta_{pu} (a_1^h +a_2^h )
+\frac32  \alpha_{3,\,{\rm EW}}^{h} 
+ \frac32 \alpha_{4,\,{\rm EW}}^{p,\,h} \right]\,,
\end{eqnarray}
where $\beta_i^h (V_1 V_2 )  = b_i^h (V_1 V_2 ) f_B f_{V_1} f_{V_2}  /A^h_{V_1 V_2 } $.
For $B^- \to K^{*-} \phi $ add the term $\delta_{pu} \beta_2^h$ to ${\cal A}^h_{\bar{B}^0 \rightarrow  K^{*0} \phi }$.
The arguments $(V_1 V_2 )$ of $a^h_i$, $\alpha^h_i$, $\beta^h_i$
have been suppressed, but it is understood that they are to be identified with 
the subscripts $(V_1 V_2 )$ of the prefactors $A_{V_1 V_2 }^h$.
The new annihilation coefficients $b^h_{i}$, and amplitudes for other decays are given in \cite{prd}. 
(To first approximation, all annihilation coefficients except $b_3^h$ can be  ignored in the above amplitudes.) $b_{1,2}^h$ arise from current-current operator annihilation graphs.
$b_4^h$ arise from QCD annihilation graphs with different flavor topology
than $b_3^h$, and $b^h_{4\,EW}$ are the analogous electroweak annihilation coefficients.

\begin{figure}[t]
\centerline{
\hbox{\hspace{0.2cm}
\includegraphics[width=4.5truecm,height=2.8truecm]{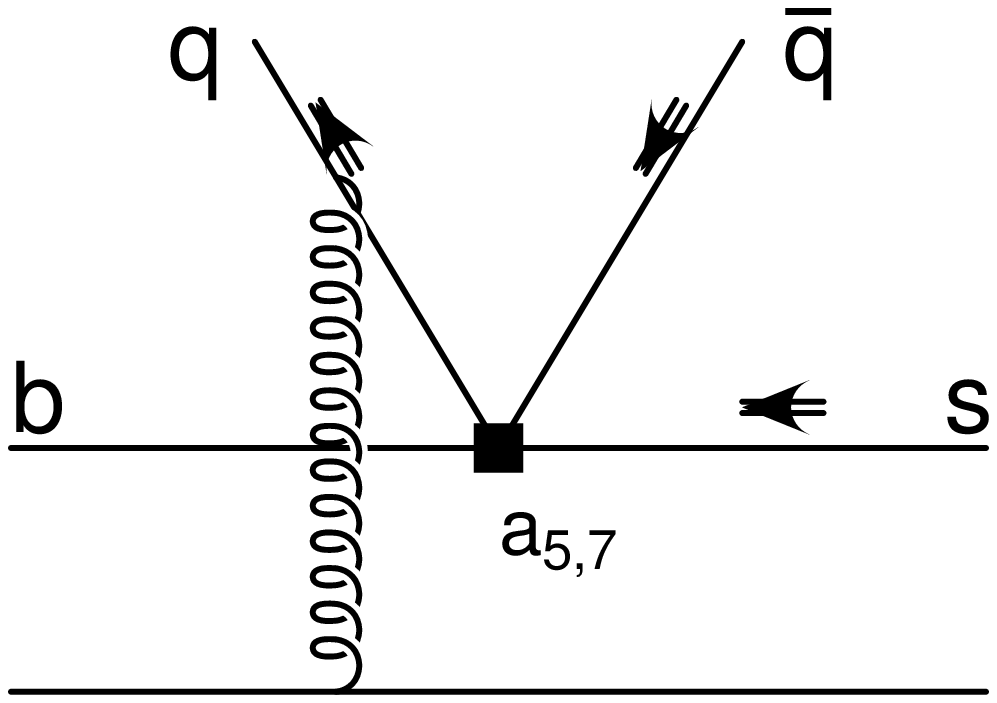}\hspace{1.5cm} 
\includegraphics[width=4.5truecm,height=2.8truecm]{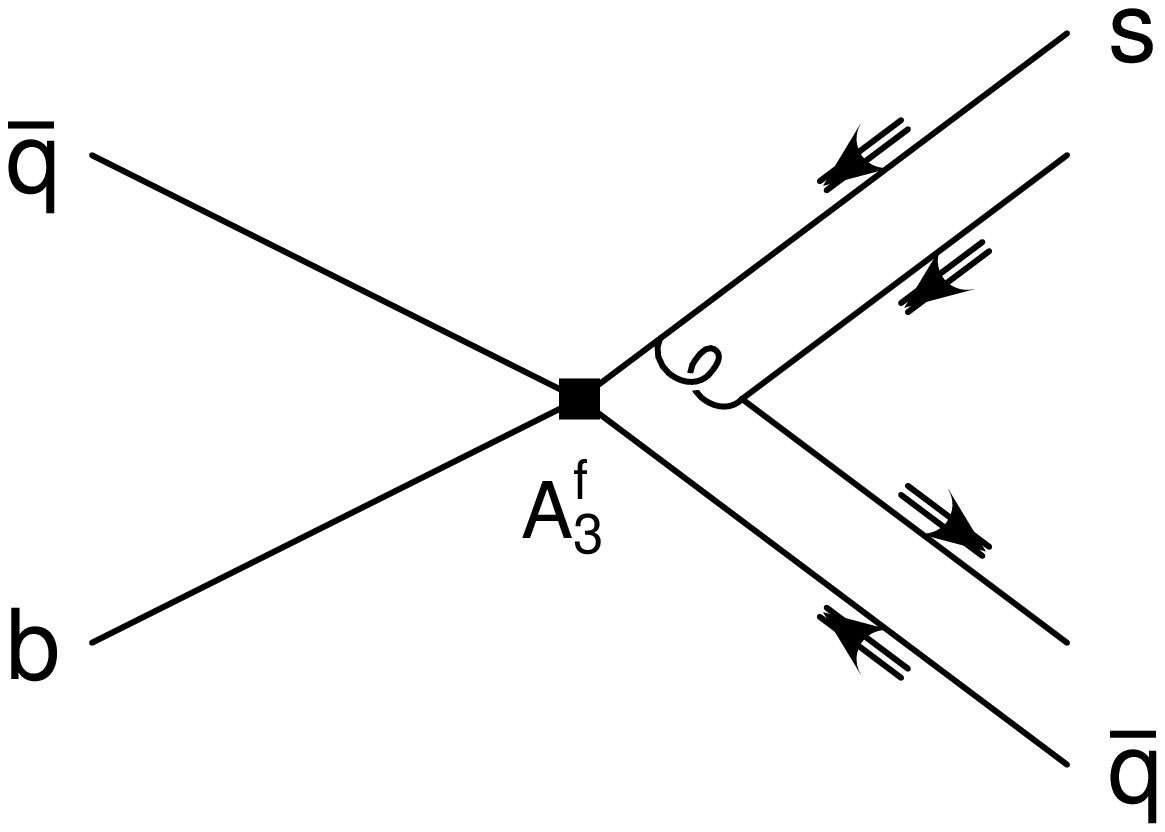}}}
\vskip -.35cm
{\caption[1]{\label{fig:penguin}
Quark helicities for $H^h_{5,7}$ (left),  $A_3^{f,\,h}$ (right).  The upward lines form $V_2$.}}
\vskip -0.2cm
\end{figure}

\section{Numerical analysis}

In our numerical analysis  the inputs 
of \cite{benekeneubert} are used, except for the form factors for which we take
$\zeta_\parallel^{K^* \,(\rho)}= .4 \pm .16\, (.31\pm .13)$ and $\zeta_\perp^{K^* \,(\rho)}=
.3 \pm .12 \,(.22 \pm .09)$.  The expanded ranges reflect the lower values
recently obtained for $\zeta^{K^*}_\perp $ from the lattice \cite{lattice2}
and from  $B\to K^* \gamma $ \cite{benekesll,aliKstgamma}.  
We also take $r^{K^* ,\, \rho}_\perp = .05 \pm .05$, spanning existing model determinations \cite{ballbraun}--\cite{BSW}. 
In the evaluation of the hard-scattering and annihilation graphs, $\alpha_s$ and the Wilson
coefficients are evaluated at an intermediated scale $\mu_h = \sqrt{\Lambda_h \mu}$,
with $\mu \in [m_b /2 , m_b]$.  The renormalization scales are varied independently
in the three classes of graphs: hard-scattering ($\mu_h^H$), annihilation ($\mu_h^A$), and those containing form factors.
The quantities $X_A$ and $X_H$ parametrizing the logarithmic divergences are varied independently for unrelated 
convolution integrals, with  $\varrho_{A,\,H}\in [0,1]$ and $\varphi_{A,\,H} \in [0,2 \pi]$.  The default values are $\varrho=\varphi=0$.

The form factor dependence can be approximately factored out of the (CP-averaged) $B \to \rho \rho$
branching ratios, yielding 
\begin{equation}
10^6 \, ( {\rm Br}_{\rho^- \rho^0 } ,\,{\rm Br}_{\rho^+ \rho^- } ) = (18.1^{+3.3}_{-2.1} ,\,27.2^{+3.6}_{-4.3} )
\times \left[\frac{|V_{ub}|}{.0037} \frac{\zeta_\parallel^\rho }{.31}\right]^2 \,.\label{rhorhoresults}
\end{equation}
The error bars include the uncertainties due to variation of all remaining inputs,
added in quadrature.
The measured values are $10^6 \,{\rm Br}^{\rm exp}_{\rho^- \rho^0} = 26.4 \pm 6.4 $
(average of \cite{BaBarRhopRho0,BelleRhopRho0}), and 
$10^6\, {\rm Br}^{\rm exp}_{\rho^+ \rho^-} = 30 \pm 4\pm 5 $ \cite{BaBarRhopRhom}.
We also obtain $f_L (\rho^- \rho^0 )= .97^{+.02}_{-.05 } \pm .01 $ and
$f_L (\rho^- \rho^+ )= .95^{+.04+.01}_{-.10-.02 }$.  The first and second sets of error bars
are due to form factor and remaining hadronic uncertainties, respectively, again added in quadrature.
The form factor errors are reduced substantially when the branching ratios are constrained to lie in their measured ranges, to approximately $^{+.02}_{-.02}$ for $\rho^- \rho^0 $ and 
$^{+.03}_{-.04}$ for $\rho^+ \rho^- $ \cite{prd}.   The relatively small polarization uncertainties are due to the absence of (in $\rho^- \rho^0$), or CKM suppression of (in $\rho^+ \rho^-$) the QCD penguin amplitudes.
The measured polarizations are 
$f^{\rm exp}_L (\rho^- \rho^0 )= .96^{+.05}_{-.07 } $ (average of  \cite{BaBarRhopRho0,BelleRhopRho0}) and $f^{\rm exp}_L (\rho^- \rho^+ )= .99\pm .03^{+.04}_{-.03}$ \cite{BaBarRhopRhom}.
Thus, the predicted longitudinal polarizations are in good agreement with experiment, and with 
naive power counting. 

\begin{figure}[t]
\centerline{
\hbox{\hspace{-.4cm}
\includegraphics[width=6.truecm]{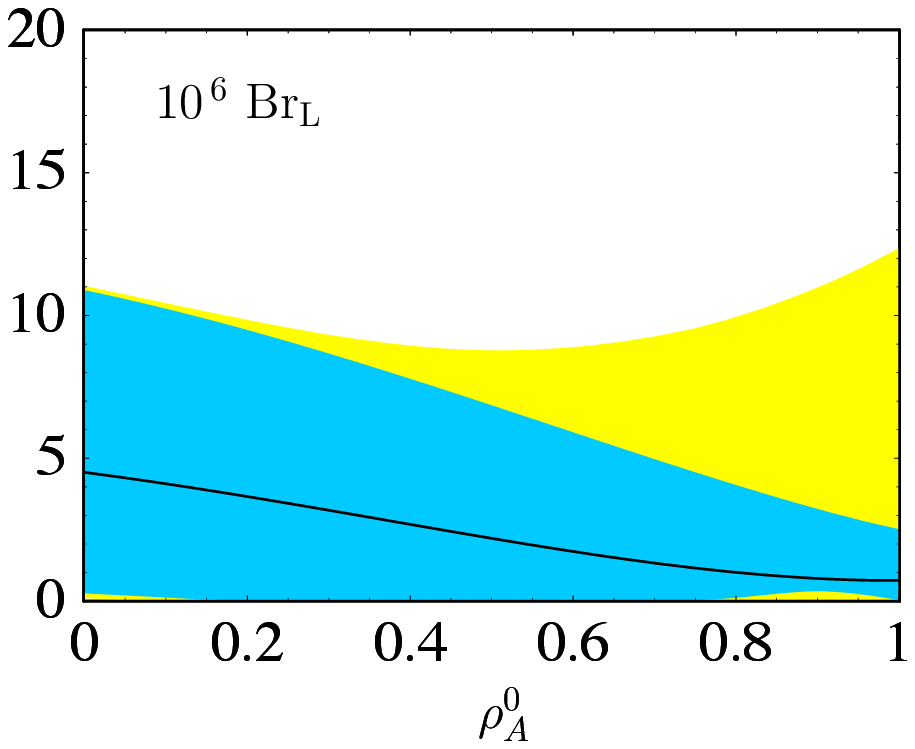}\hspace{.5cm} 
\includegraphics[width=6.truecm]{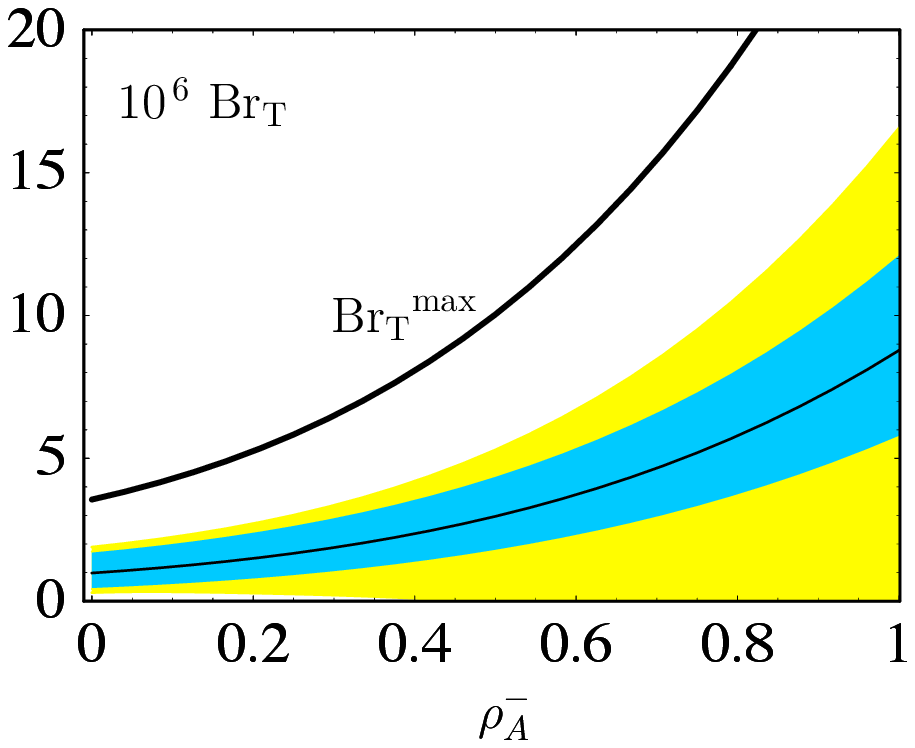}}}
\vskip -.35cm
{\caption[1]{\label{fig:penguin}
${\rm Br}_{L} ( \phi K^{*0})$ vs.  $\rho^{0}_A $ (left), ${\rm Br}_{T} ( \phi K^{*0})$ vs.  $\rho^{-}_A $ (right).
Black lines: default inputs. Blue bands:
input variation uncertainties.  Yellow bands: additional uncertainties from 
logarithmically divergent power corrections.  Thick line: ${\rm Br}^{\rm max}_T$, see text.}}
\vskip -0.2cm
\end{figure}

Averaging the Belle and BaBar $\bar B\to \phi K^{*0}$ (CP-averaged) measurements \cite{BellePhiKst,BaBarPhiKst,BaBarRhopRho0} yields 
$f^{\rm exp}_L=.49\pm .06$ and $10^6\, {\rm Br }^{\rm exp}=10.61\pm1.21$, or 
 $10^6\,{\rm Br}^{\rm exp}_L = 5.18 \pm .86$ and $10^6\, {\rm Br}^{\rm exp}_T = 5.43 \pm .88$,
 where ${\rm Br}_T = {\rm Br}_\perp + {\rm Br}_\parallel $ is the total transverse branching ratio.
Without annihilation we obtain
$10^6\, {\rm Br}_L = 5.2^{+6.8+.9}_{-4.7-.8} $
and $10^6 \,{\rm Br}_T = .6^{+.6+.4}_{-.4-.3} $,
where the second (first) set of error bars is due to variations of $X_H$ (all other inputs).  In Figure 4, ${\rm Br}_L$ and 
${\rm Br}_T$ are plotted versus the annihilation parameters $\rho^0_A$ and $\rho^-_A$
entering $A_3^{f ,0}$ and $A_3^{f ,-}$, respectively.
The black curves are obtained for central values of  all inputs, with
default values for all annihilation and hard spectator interaction parameters other than 
$\rho^0_A ,\,\rho^-_A$.
The blue bands are obtained by adding the uncertainties due to variations of the inputs in quadrature, keeping
default annihilation and hard spectator interaction parameters.  
The widths of the bands are dominated by the form factor uncertainties.
The yellow bands also include, in quadrature,
the uncertainties due to variations of all $\varrho_{H,\,A}, \varphi_{H,\,A}$ and $\mu_h^A$.
The thick curve gives the maximum values obtained for ${\rm Br}_T$ under
simultaneous variation of all inputs.  The absolute branching ratios suffer from  
large theoretical uncertainties, as is usually the case.
Nevertheless, it is clear that the contributions of $A_3^{f ,0}$ and $A_3^{f ,-}$ to the QCD annihilation amplitudes can be 
${\cal O}(1)$ numerically even though they are formally ${\cal O}(1/m^2)$.
This can be traced to the quadratic dependence on the divergences ($X^2_A$)
and the large coefficient $N_c C_6$ in $b_3^{f,\,h}$.
The quantities $X_A^0 $ and $X_A^-$, as well as the renormalization scales and 
form factors entering $\bar{\cal A}^0 $ and $\bar{\cal A}^-$ are, a-priori, unrelated.  
Figure 4 therefore implies that 
the measurements of ${\rm Br}_L $ and ${\rm Br}_T$ can easily be accounted for simultaneously.
According to Figure 4, $f_L (\phi K^{*-} ) \approx 50\%$ can also be accounted for given that
the $\phi K^{*-}$ and $\phi K^{*0}$ amplitudes only differ by a small current-current operator annihilation graph.
\vspace{0.5cm}
\newline\noindent{\it A test for right-handed currents}
\vspace{0.2cm}

In Figure 5 (left) the predicted ranges for $f_\perp /f_\parallel $ and ${\rm Br}_T$ 
are studied simultaneaously for $\bar B \to \phi K^{*0}$ in the Standard Model.
The `default{'} curve is again obtained by varying $\varrho_A^-$ in the range $[0,1]$,
keeping all other inputs at their default values, and the error bands are obtained by adding uncertainties in quadrature as in Figure 4.
Evidently, the 
second relation in (\ref{SMpred}) holds at next-to-leading order, particularly
at larger values of ${\rm Br}_T$ where 
QCD annihilation dominates ${\rm Br}_\perp $ and ${\rm Br}_\parallel$.
We also plot the maximum values attained for $f_\perp /f_\parallel $ under 
simultaneous variation of all inputs.
The result is sensitive to $r_\perp$, 
as it largely determines the relative signs and magnitudes
of the  `form factor{'} terms in $\bar {\cal A}^-$ and $\bar{\cal A}^+ $, see (\ref{Av1v2SCET}).
The thick black curve (corresponding to ${\rm Br}_T^{\rm max}$ in Figure 4)
and blue curve give maxima for  
$r_\perp \ge 0$, in accord with existing model determinations,
and $r_\perp \ge -.10$, respectively.
A ratio in excess of the Standard Model range, e.g.,  
$f_\perp /f_\parallel > 1.5$ if $r_\perp > 0$,
would signal the presence of new right-handed currents.
We mention that non-vanishing
CP-violating triple products in pure penguin decays like $\bar B \to \phi K^{*}$ would not be a signal for right-handed currents if  
significant strong phase differences ($\ne 0\, {\rm mod}\, \pi$) existed between 
${\cal \bar A}_{0,\parallel} $ and ${\cal \bar A}_\perp$ \cite{valencia,dattalondon}.
There is some experimental indication for such phase differences \cite{BaBarPhiKst},
which is to be expected if annihilation amplitudes are important.  

\begin{figure}[t]
\centerline{
\hbox{\hspace{-.4cm}
\includegraphics[width=6.truecm]{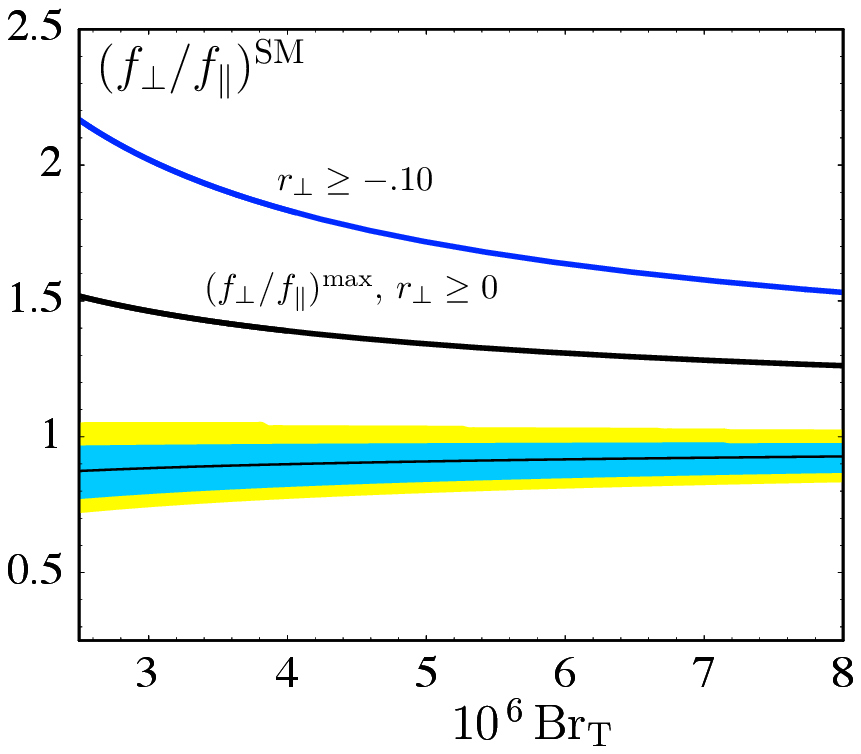}\hspace{.5cm} 
\includegraphics[width=6.truecm]{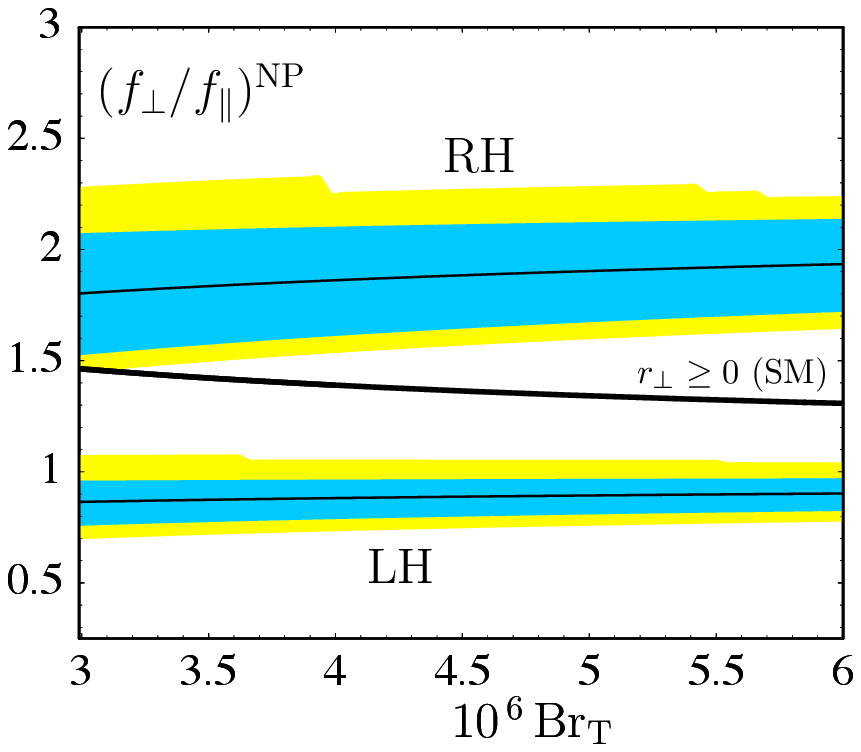}}}
\vskip -.35cm
{\caption[1]{\label{fig:penguin}
$f_\perp /f_\parallel $ vs. ${\rm Br}_T$ in the SM (left), and with new
RH or LH currents (right).  Black lines, blue bands, and yellow bands are as in Figure 4.  Thick lines: $(f_\perp /f_\parallel)^{\rm max}$ in the Standard Model for indicated ranges of $r_\perp$.}}
\vskip -0.2cm
\end{figure}

Right-handed currents are conventionally associated with effective operators
$\tilde{Q}_i$, obtained from the Standard Model operators $Q_i$ 
by interchanging $V\!-\!A \leftrightarrow V\!+\!A$.
The final states in ${\cal A}_{0,\parallel}$ (${\cal A}_\perp$) are parity-even (parity-odd), 
so that the  i'th pair of Wilson coefficients enters as \cite{oldertalks}
\begin{equation}\label{NPamps}
{\cal A}_{0,\parallel } \propto C_i^{\rm SM} \!+\! C_{i}^{\rm NP} \!-\! \tilde{C}_{i}^{\rm NP},
~~{\cal  A}_{\perp} \propto C_i^{\rm SM} \!+\! C_{i}^{\rm NP}\! +\! \tilde{C}_{i}^{\rm NP}.
\end{equation}
The different combinations allow for large modifications to (\ref{SMpred}).
$f_L$ suffers from prohibitively large theoretical uncertainties.
However, $f_\perp /f_\parallel $ is a much cleaner observable.
For illustration, new contributions to the QCD penguin operators are considered in Figure 5 (right).
At the New Physics matching scale $M$, these can be parametrized as 
${\stackrel{\scriptscriptstyle{(\sim)}}{C_4}} = {\stackrel{\scriptscriptstyle{(\sim)}}{C_6}} = - 3\, {\stackrel{\scriptscriptstyle{(\sim)}}{C_5}} = -3\, {\stackrel{\scriptscriptstyle{(\sim)}}{C_3}}={\stackrel{\scriptscriptstyle{(\sim)}}{\kappa}} $.
For simplicity, we take $M\!\approx\! M_W$ and consider two cases:
$\kappa = -.007 $ (lower bands) or $\tilde{\kappa}=-.007$ (upper bands),
corresponding to $C^{NP}_{4\, (5)} (m_b)$ or $\tilde{C}^{NP}_{4 \,(5)} (m_b) \approx  .18\,  C_{4\, (5)}^{SM}  (m_b)$, and
$C^{NP}_{6 \,(3)} (m_b)$ or
$\tilde{C}^{NP}_{6\, (3)} (m_b) \approx  .25  \,C_{6 \,(3)}^{SM}  (m_b)$, respectively.
The default curves and error bands are obtained as in the Standard Model case.   Clearly,
moderately sized right-handed currents could increase $f_\perp /f_\parallel$ well beyond the Standard Model
range if $r_\perp \ge 0$.
However, new left-handed currents would have little effect.
\newpage
\noindent{\it Dipole operators versus four-quark operators}
\vspace{0.2cm}

The suppression of dipole operator effects in the transverse modes
has important implications for New Physics searches.  
For example, in pure penguin decays to CP-conjugate final states $f$, e.g.,
$\bar B\to \phi\, ( K^{*0} \to  K_s \pi^0 )$, if the transversity basis time-dependent CP asymmetry parameters $(S_f)_{\perp} $ and $(S_f)_{\parallel} $ are consistent with  $(\sin 2 \beta)_{J/\psi K_s}$, and $(S_{f})_0$ is not,
then this would signal new CP violating contributions to the
chromomagnetic dipole operators.  However, deviations in $(S_{f})_\perp $ or
$(S_{f})_\parallel $ would signal new CP violating four-quark operator contributions.  
If the
triple-products $A_T^0 $ and  $A_T^\parallel $  \cite{valencia,dattalondon}
do not vanish and vanish, respectively, 
in pure penguin decays, then this
would also signal new CP violating contributions to the
chromomagnetic dipole operators.  (This assumes that a significant strong phase difference is measured between ${\cal \bar A}_\parallel $ and ${\cal \bar A}_\perp$.)
However, non-vanishing $A_T^\parallel $, or non-vanishing transverse direct CP asymmetries
would signal the intervention of four-quark operators.
The above would help to discriminate between different explanations for 
an anomalous time-dependent CP asymmetry in $B \to \phi K_s$, i.e., $S_{\phi K_s}$, which fall broadly into two categories:  
radiatively generated dipole operators, 
e.g., supersymmetric loops; or tree-level four-quark operators, e.g., 
flavor changing (leptophobic) $Z^\prime $ exchange \cite{berger}, $R$-parity violating couplings \cite{datta}, or color-octet exchange \cite{burdmann}.  Finally, a large $f_\perp /f_\parallel $ would be a signal for right-handed {\it four-quark} operators.
\vspace{0.5cm}
\newline\noindent{\it $SU(3)_F$ violation and $B\to \rho K^*$}
\vspace{0.2cm}

We have seen that the large transverse $\phi K^* $ polarization can be accounted for in the Standard Model via the QCD penguin annihilation graphs.  
Would this
necessarily imply large transverse $\rho K^*$ polarizations? 
To answer this question we need to address $SU(3)_F $ flavor symmetry breaking in annihilation.
For simplicity, we have thus far estimated the annihilation amplitudes using asymptotic light-cone distribution amplitudes \cite{BBNS,benekeneubert}.
However, for light mesons containing a single strange quark,
non-asymptotic effects should shift the weighting of the distribution amplitudes towards larger strange quark momenta.  $SU(3)_F$ violation in processes involving $s \bar s$ popping versus light quark popping, e.g., annihilation, can therefore be much larger than the 
canonical 20\%, due to the appearance of inverse moments of the distribution amplitudes \cite{pirjolstewart}.
This can account for the order of magnitude hierarchy
between the $\bar B \to D^0 \pi^0 $ and $\bar B \to D_s^+ K^- $ rates \cite{pirjolstewart}.  Similar considerations may also explain the ${\cal O}(1)$ flavor violation empirically observed in high energy $e^+ e^- $ fragmentation, e.g., in kaon versus pion multiplicities, or $K^*$ versus $\rho$ multiplicities at the $Z$.  In particular, the relative 
probability for $s\bar s $ popping versus  
$u\bar u$ or $d \bar d$ popping in JETSET fragmentation Monte Carlo's must be tuned to $\approx .3$ \cite{JETSET}.

\begin{figure}[t]
\centerline{
\hbox{\hspace{-0.4cm}
\includegraphics[width=6.0truecm]{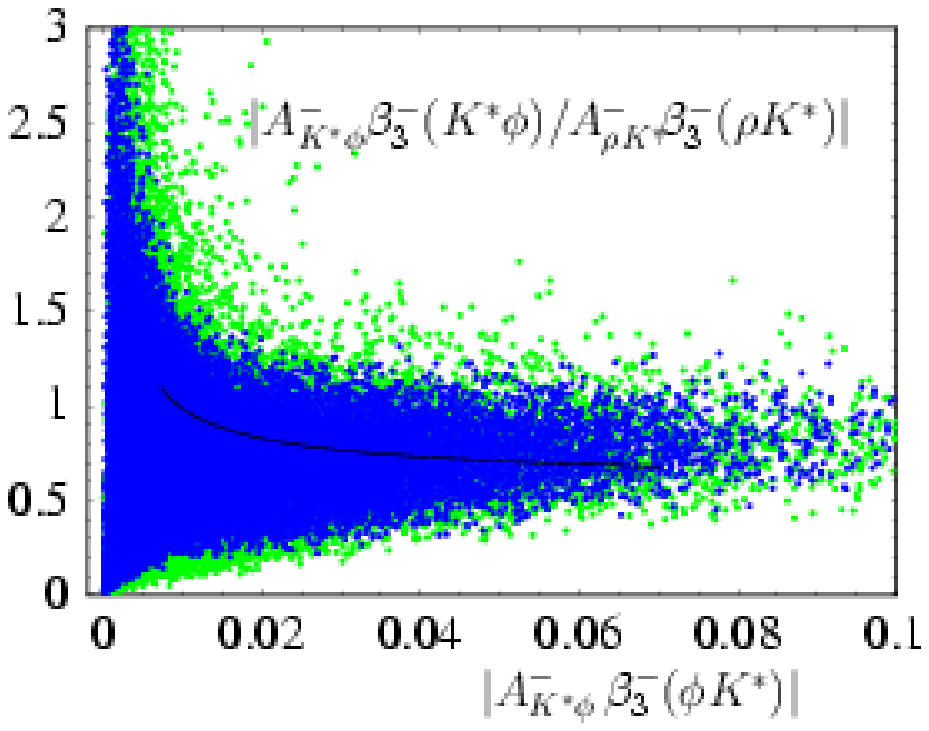}\hspace{.7cm} 
\includegraphics[width=6.0truecm]{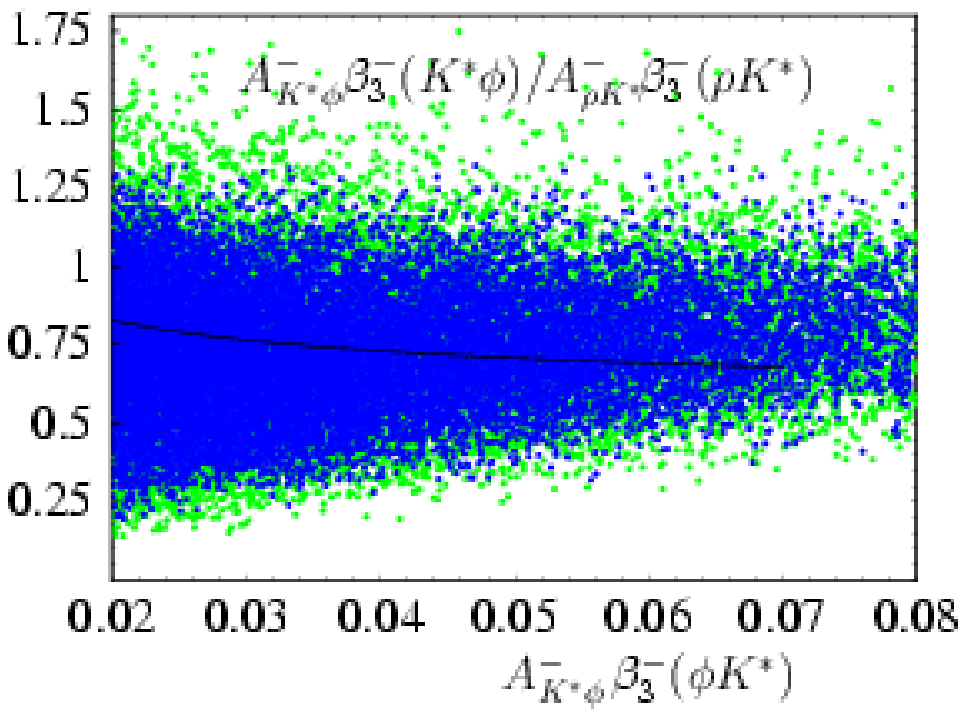}}}
\vskip -.35cm
{\caption[1]{\label{fig:SU3}
Scatter plots of $|A_{K^* \phi}^- \beta_3^- (\phi K^* ) /A_{\rho K^* }^- \beta_3^- (\rho K^* ) |$ vs. $|A_{K^* \phi}^- \beta_3^- (\phi K^* ) | $ for $\varphi_A \ne 0$, and $A_{K^* \phi}^- \beta_3^- (\phi K^* ) /A_{\rho K^* }^- \beta_3^- (\rho K^* ) $ vs. $A_{K^* \phi}^- \beta_3^- (\phi K^* )  $ for $\varphi_A =0$,
for full (green), halved (blue) and default (solid line) ranges of Gegenbauer moments, see text.}}
\vskip -0.2cm
\end{figure}

The dominant $B \to VV$ QCD annihilation amplitudes 
$A_3^{f,\, -}$ and $A_3^{f,\,0}$ involve products of inverse moments, see
(\ref{A3f}).  
The $SU(3)_F$ violation discussed above can be estimated by including the second and third terms in the Gegenbauer expansions for the distribution amplitudes.  
The first Gegenbauer moments $\alpha_{1,\,\perp}$, $\alpha_{1,\,\parallel}$ determine the asymmetries of the corresponding leading-twist distribution amplitudes, i.e., the inverse moments of $\phi_{\perp}$ are given by $\langle \bar u^{-1}   \rangle_{\perp },\, \langle u^{-1}   \rangle_{\perp }= 3( 1\pm \alpha_{1, \perp} + \alpha_{2,\perp } )$, and similarly
for $\phi_\parallel$.  Note that the
first moments vanish for the symmetric $\phi $ and $\rho$ mesons.
For illustration, two sets of intervals for the moments are considered: 
$\alpha_{1,\perp (\parallel)}^{K^*} = .2\pm .2$, $\alpha_{2,\perp (\parallel)}^{K^* ,\rho}=.1\pm .3$,
$\alpha_{2,\perp (\parallel)}^\phi = 0\pm .3$, as in \cite{benekeneubert}; and a more restrictive set
with same central values but halved intervals. 
$\alpha_{1,\,\perp (\parallel)}^{K^*} > 0$ is required, since 
the $s$-quark should carry the larger fraction of the $K^*$ light-cone momentum.
Similarly, we require that the light-cone momentum fraction of the $s$-quark (light antiquark)
in the $K^*$ is greater than (less than) that of the quark (antiquark) in the $\rho$ and $\phi$
or, equivalently, 
\begin{equation}
\label{constraint}
\langle\bar u^{-1}\rangle^{K^*}_{\perp (\parallel)}  > \langle\bar u^{-1} \rangle^{\rho,\phi}_{\perp (\parallel)},~~~~\langle u^{-1} \rangle^{K^*}_{\perp (\parallel)}  <   \langle u^{-1} \rangle^{\rho,\phi}_{\perp (\parallel)}\,,
\end{equation}
which imposes the constraints
$\alpha^{K^*}_{1,\perp (\parallel)} > |\alpha_{2,\perp (\parallel)}^{K^*} -\alpha_{2,\perp (\parallel)}^{\rho,\phi} |$.
The logarithmic divergences in the inverse moments are parametrized as before.  For simplicity, the $X_A$
are taken equal and independent of the final state,  with $\varrho_A \le 1$ and $\varphi_A \in [0,2\pi]$.   In \cite{benekeSU3}, $SU(3)_F$ violation was studied with asymptotic distribution amplitudes by varying
the $X_A$.

The scatter plots in Figure 6 illustrate $SU(3)_F$ violation in
the QCD penguin annihilation amplitudes $A_{K^* \phi }^- \, \beta_3^- (\phi K^* )$ and
$A_{\rho K^*}^- \,\beta_3^- (\rho K^* ) $, see (\ref{VVamplitudes}).
For simplicity, only the contributions of $A_3^{f,\,-}$ are included (to good approximation,
the other terms in $b_3^-$ can be neglected). 
Two cases are shown: arbitrary strong phase $\varphi_A  \in [0,2 \pi]$; and vanishing strong phase $\varphi_A=0 $.
The Gegenbauer moments are sampled in the intervals given above, subject to the constraint
(\ref{constraint}); $\mu_h^A$ lies in the usual range, $\rho_A \in [0,1]$,
and the remaining inputs are set to their default values.
For comparison, the default non-annihilation $ \phi K^{*0}$ amplitude (in units of $-i \lambda_c^{(s)}{G_F/\sqrt{2}}$) is $\approx .026 $, with negligible strong phase.  The total
negative helicity $\bar B^0 \to \phi K^{*0}$ amplitude observed is about a factor of three larger, corresponding to $A_{K^* \phi}^- \beta_3^- (K^* \phi ) \sim .05$
and, according to Figure 6, $ .9  \ltap A_{\rho K^* }^- \beta_3^- (\rho K^* ) /A_{K^* \phi }^-  \beta_3^- (\phi K^* ) \ltap  2.5 $.
We therefore expect $f_L (K^{*0}  \rho^\pm) \ltap f_L (\phi K^{*0})$ in the Standard Model.   
The other $\rho K^* $ modes, containing 
`tree-level' amplitudes, will be discussed in \cite{prd}.
BaBar has measured $f_L (\rho^0 K^{*+}) = 0.96^{+0.04+0.04}_{-0.15-0.04}$
\cite{BaBarRhopRho0}.  Given the large errors, this
is still consistent with the low $\phi K^*$  longitudinal polarization.

\section{Conclusion}

We have presented an analysis of polarization in $B $ decays to light vector meson pairs
beyond naive factorization, using QCD factorization.  Formally, the longitudinal polarization satisfies $1-f_L = {\cal O}(1/m^2)$, as in naive factorization.  However, we saw that the contributions of a particular QCD penguin annihilation graph
which is formally ${\cal O}(1/m^2 )$ can be ${\cal O}(1)$ numerically in longitudinal and 
negative helicity $\Delta S\!=\!1$ $\bar B$ decays.  Consequently, the observation of $f_L (\phi K^{*0,-} ) \approx 50\%$ can be accounted for in the Standard Model, with large theoretical errors. 
However, $f_L (\rho^+ \rho^0 )$ and $f_L (\rho^+ \rho^- )$ are predicted to be close to 1 with small 
theoretical errors, 
in agreement with observation.
We have shown that the ratio of transverse rates in the transversity basis satisfies
$\Gamma_\perp /\Gamma_\parallel = 1 + {\cal O}(1/m )$, in agreement with naive power counting.
A ratio in excess of the predicted Standard Model range would signal the presence of new right-handed currents in dimension-6 four-quark operators.  
The maximum ratio attainable in the Standard Model is sensitive to the $B \to V$ form factor combination $(1+m_{V}/{m_B} ) 
{ A_1}  - (1- {m_{V}}/{m_B} ) { V} $ or $r_\perp$, see (\ref{rperp}), which controls helicity suppression in form factor transitions.  Existing model determinations give a positive sign for $r_\perp$, which would imply
$\Gamma_\perp (\phi K^{*} ) /\Gamma_\parallel (\phi K^{*} )  < 1.5 $ in the Standard Model.
However, the maximum would increase for negative values.
The magnitude and especially the sign of $r_\perp $ is an important issue which needs to be clarified with dedicated lattice studies.

The contributions of the $b \to sg$ dipole operators to the transverse modes were found to be highly suppressed, due to angular momentum conservation.
Comparison of CP-violation 
involving the longitudinal modes with CP-violation only involving the transverse modes, in pure penguin $\Delta S=1$ decays, could therefore distinguish between new contributions to the dipole  and four-quark operators.
More broadly, this could distinguish between scenarios in which New Physics effects are loop induced and scenarios in which they are tree-level induced, as it is difficult to obtain ${\cal O}(1)$ CP-violating effects from dimension-6 operators beyond tree-level.  

We have seen that the asymmetry of the $K^{(*)}$ meson light-cone distributions
generically leads to ${\cal O}(1)$ $SU(3)_F$ flavor symmetry violation in annihilation amplitudes, as pointed out in \cite{pirjolstewart}.  In particular,  $s \bar s$ popping can be substantially
suppressed relative to light quark popping.
This implies that the longitudinal polarizations should satisfy $f_L ( \rho^\pm K^{*0}) \ltap f_L (\phi K^{*})$ in the Standard Model.
Consequently, $f_L ( \rho^\pm K^{*0}) \approx 1$ would indicate 
that $U$-spin violating New Physics entering mainly in the $b\to s \bar s s$ channel 
is at least partially responsible for the small values of $f_L (\phi K^{*} )$.
One possibility would be right-handed vector currents; they could interfere constructively (destructively) in the perpendicular (longitudinal and parallel) transversity amplitudes.  Alternatively, a parity symmetric realization
would only affect, and increase the perpendicular amplitude \cite{oldertalks}. 
Either case would lead to $\Gamma_\perp  > \Gamma_\parallel$, 
and could thus be ruled out.
A more exotic possibility is tensor currents;  they would contribute to the longitudinal and transverse amplitudes at subleading and 
leading power, respectively.  If left-handed, i.e., of the form
$\bar s \sigma_{\mu \nu} (1+\gamma_5 ) b\, \bar s \sigma^{\mu \nu} (1\pm \gamma_5 ) s$, then $\Gamma_\perp \approx \Gamma_\parallel $ would be maintained.  
Finally, ${\cal O}(1)$ $SU(3)_F$ violation is possible in all QCD penguin amplitudes,
given that the annihilation topology components can be comparable to, or greater than the penguin topology components.
This is especially true of decays to $VV$
and $VP$ final states which, unlike decays to $PP$ final states, do not receive large contributions from $(S-P) (S+P)$ chirality
penguin topology matrix elements. 
Certain applications of $SU(3)_F$ symmetry in $B$ decays should therefore be reexamined.

\vspace{0.15cm}  
{\it Acknowledgments:} 
This work originated during the 2003 Summer Workshop on Flavor Physics at the Aspen 
Center for Physics.  I would like to thank Martin Beneke, Alakhabba Datta, Keith Ellis, Andrei Gritsan, Yuval Grossman, Dan Pirjol, and Ian Stewart for useful conversations.  I am especially grateful to Matthias Neubert for discussions of many points addressed in this paper.  This work was supported by the Department of Energy under Grant DE-FG02-84ER40153.

\end{document}